\documentclass[sigconf,authorversion,nonacm]{acmart}

\AtBeginDocument{%
  }

\makeatletter
\newcommand\asterfootmark{%
  \begingroup
    \renewcommand\thefootnote{\fnsymbol{footnote}}%
    \footnotemark[1]%
  \endgroup
}
\newcommand\asterfoottext{%
  \begingroup
    \renewcommand\thefootnote{\fnsymbol{footnote}}%
    \footnotetext[1]{Work done during internship at NVIDIA}%
  \endgroup
}
\makeatother

\makeatletter
  \renewcommand\@printendtopmatter{
  \par\medskip
  }
\makeatother

\usepackage{microtype}
\usepackage{graphicx}
\usepackage{subfigure}
\usepackage{booktabs} 
\usepackage{multicol}
\usepackage{multirow}
\usepackage{siunitx}
\usepackage{xcolor,colortbl}

\newcommand\hc{ \rowcolor{teal!15}}

\newcommand{\expect}[2]{\mathbb{E}_{#1}\left[ #2 \right]}

\usepackage{hyperref}

\usepackage{amsmath} 
\usepackage{amsfonts}

\begin{document}

\title{FGMP: Fine-Grained Mixed-Precision Weight and Activation Quantization for Hardware-Accelerated LLM Inference}

\author{%
  Coleman Hooper\textsuperscript{1,2}\asterfootmark,\enspace
  Charbel Sakr\textsuperscript{1},\enspace
  Ben Keller\textsuperscript{1},\enspace
  Rangharajan Venkatesan\textsuperscript{1},\enspace \\
  Kurt Keutzer\textsuperscript{2},\enspace
  Sophia Shao\textsuperscript{2},\enspace
  Brucek Khailany\textsuperscript{1}%
}
\affiliation{%
  \institution{%
    \textsuperscript{1}NVIDIA,
    \textsuperscript{2}University of California, Berkeley
  }
  \country{}
}

\begin{abstract}

Quantization is a powerful tool to improve large language model (LLM) inference efficiency by utilizing more energy-efficient low-precision datapaths and reducing memory footprint.
However, accurately quantizing LLM weights and activations to low precision is challenging without degrading model accuracy.
We propose fine-grained mixed precision (FGMP) quantization, a post-training mixed-precision quantization methodology that maintains accuracy while quantizing the majority of weights and activations to reduced precision.
We co-design hardware which exploits the mixed-precision data representation to achieve greater energy efficiency and which quantizes activations to mixed precision on the fly.
Our work makes the following contributions:
1) We develop a policy that uses the perturbation in each value, weighted by the Fisher information, to select which weight and activation blocks to keep in higher precision.
This approach preserves accuracy by identifying which weight and activation blocks need to be retained in higher precision to minimize the perturbation in the model loss.
2) We also propose a sensitivity-weighted clipping approach for fine-grained quantization which helps retain accuracy for blocks that are quantized to low precision.
3) We then propose hardware augmentations to leverage the efficiency benefits of FGMP quantization.
Our hardware implementation encompasses i) datapath support for FGMP at block granularity, and ii) a mixed-precision activation quantization unit to assign activation blocks to high or low precision on the fly with minimal runtime and energy overhead.
Our design, prototyped using NVFP4 (an FP4 format with microscaling) as the low-precision datatype and FP8 as the high-precision datatype, facilitates efficient FGMP quantization, attaining $<$1\% perplexity degradation on Wikitext-103 for the Llama-2-7B model relative to an all-FP8 baseline design while consuming 14\% less energy during inference and requiring 30\% less weight memory.
\end{abstract}
\maketitle
\asterfoottext
\pagestyle{plain}

\section{Introduction}

\begin{figure}[t]
\centering
\includegraphics[width=\linewidth]{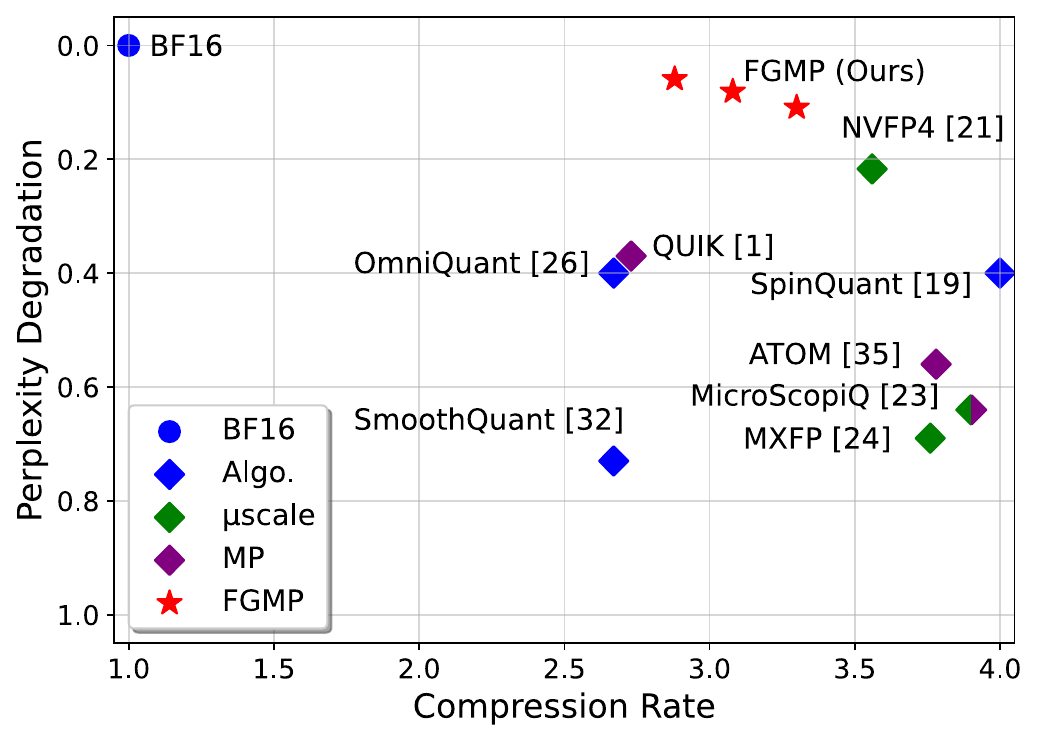}
\vspace*{-5mm}
 \caption{
Perplexity degradation (less is better) versus compression rate (higher is better), evaluated using Llama-2-7B on Wikitext-103 with 4-bit weight-activation quantization. 
Compression rate is computed as 16 divided by the average bit width for weights and activations (assuming 4K context length).
We group prior work into algorithm-only approaches that use integer quantization (``Algo.''), prior work on microscaling quantization (``$\mu$scale''), and prior methods for mixed-precision quantization (``MP''). Our method, FGMP, (shown here with 70\%, 80\%, and 90\% of blocks in FP4) attains reduced perplexity degradation relative to existing post-training quantization methods.\protect\footnotemark
}
  \label{fig:ppl-vs-compression}
  \vspace{-6mm}
\end{figure}

\begin{figure*}[t]
\centering
\includegraphics[width=\linewidth]{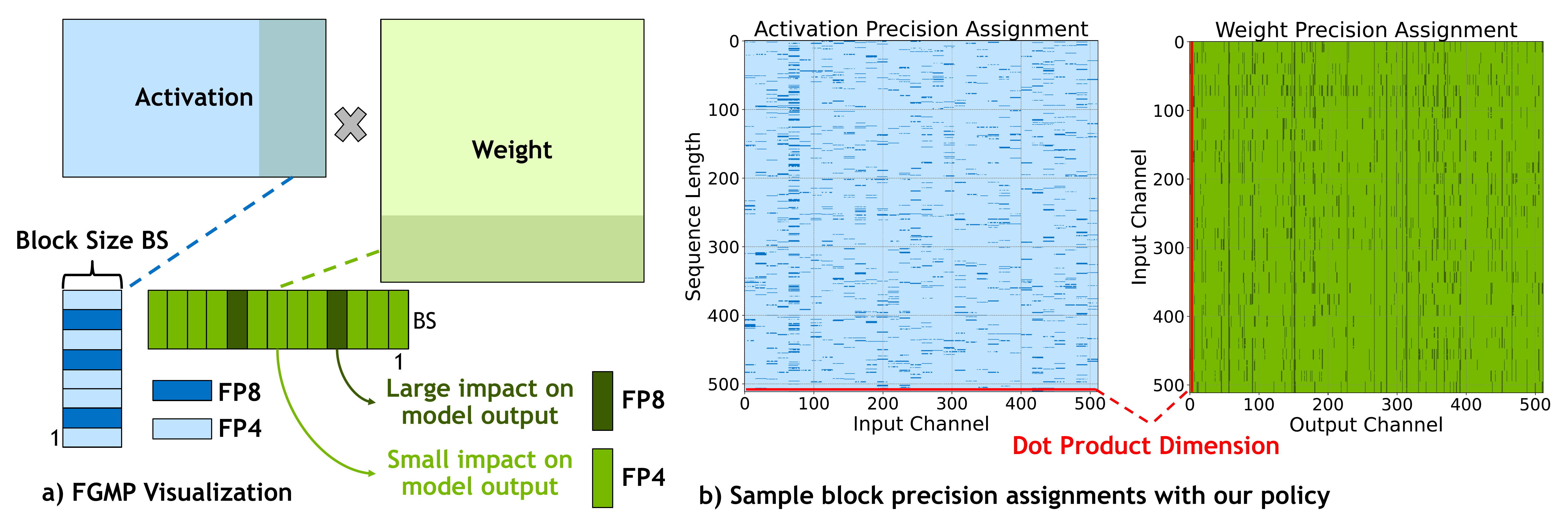}
\vspace*{-5mm}
 \caption{
a) Diagram outlining our approach for FGMP quantization at block granularity. 
We quantize both weights and activations to mixed precision, preserving sensitive blocks in higher precision (FP8) and quantizing the rest to low precision (NVFP4).
b) Visualization of a slice of the activation and weight precision assignments for Layer 7 Fully Connected 1  (``FC1'') in the Llama-2-7B model using our policy with 10\% of blocks in FP8 and with our sensitivity-weighted clipping method applied.
The NVFP4 and FP8 blocks are interleaved in an unstructured manner.
}
  \label{fig:main}
  \vspace{-5mm}
\end{figure*}

Large language models (LLMs) are scaling rapidly in both model size and computational requirements, which makes it challenging to efficiently perform inference using these networks. 
One approach for reducing both computational and memory requirements for LLM inference is quantization of model weights and activations to low-precision number formats.
However, aggressive quantization of all weights and activations is challenging due to the presence of data particularly sensitive to perturbations, such as large-magnitude outliers in weights and activations.
Quantization-aware training (QAT) may be employed to mitigate these perturbations \cite{jacob2018quantization,nagel2021white}; however, the costs of re-training and unavailability of training data with recent LLMs can make QAT difficult and motivates post-training quantization (PTQ) approaches \cite{frantar2022gptq,lin2024awq}.
In this paper, we explore techniques to improve PTQ inference accuracy.

\footnotetext{Note that 6-bit configurations are reported for \cite{shao2023omniquant,xiao2023smoothquant} since the corresponding W4A4 configurations have greater than 1 PPL loss, and \cite{shao2023omniquant,zhao2024atom} quantize the KV cache activations in addition to the linear layers.}

Using low precision (e.g. 4-bit) quantization for weights and activations for LLMs is desirable since it allows the use of energy-efficient low-precision compute datapaths (in addition to memory savings). 
However, accurate post-training quantization with weights and activations in 4-bit precision is challenging; prior works still have substantial accuracy degradation when quantizing to low precision for both weights and activations \cite{xiao2023smoothquant,liu2024spinquant,shao2023omniquant,ashkboos2024quarot}.
The reason that these methods struggle to retain accuracy in aggressive regimes like 4-bit quantization is due to challenges with numerical outliers which skew the quantization range \cite{kim2023squeezellm, dettmers2022llm,dettmers2023spqr}, as well as disproportionately sensitive values \cite{kim2023squeezellm}. 
To address this challenge, previous works have applied \textit{mixed-precision} quantization, where a mixture of high-precision and low-precision quantization is used to preserve accuracy by retaining particularly sensitive or hard to quantize parts of the network in higher precision \cite{zhao2024atom,ashkboos2023towards}.
However, existing methods typically perform coarse-grained mixed precision (e.g., at the granularity of rows of a matrix \cite{zhao2024atom,ashkboos2023towards}).
As oulined in Figure \ref{fig:main}, the distribution of sensitive and hard-to-quantize values is unstructured in both LLM weights and activations; existing coarse-grained mixed-precision quantization methods cannot fully adapt to these unstructured distributions. 

Our work addresses the challenges with existing mixed-precision quantization methods by leveraging fine-grained mixed-precision (FGMP) quantization, in which we identify small, contiguous regions of hard-to-quantize weights and activations to be retained in a higher-precision data format while the majority of weights and activations are quantized to a lower-precision format.
We pursue mixed-precision quantization at much finer \textit{block} granularity.
We define a \textit{block} as a small ($<$100 elements) 1-dimensional vector along the dot product dimension of each tensor (as in \cite{huang2024slim}).
We perform FGMP quantization by independently assigning each block to either a higher-precision or lower-precision data format (see Figure~\ref{fig:main}).
This provides improved accuracy since it allows for more flexible precision assignment, adapting to different distributions of sensitive values (both inter-layer and intra-layer).
However, although FGMP is a promising approach for maintaining model accuracy with the majority of values in reduced precision, it is challenging to leverage the potential efficiency benefits due to the difficulty of performing computation using different precisions for adjacent blocks.
There are also distinct challenges with applying FGMP quantization to activations, as we need to be able to efficiently identify hard-to-quantize activation blocks online during inference to quantize the output activations of each layer.
Even if we leveraged low-precision datapaths for the majority of computations, software implementations for dynamically selecting precisions when quantizing blocks of activations may be impractical.

Our work aims to address these challenges through hardware-software co-design in order to enable fine-grained mixed-precision quantization while supporting low-precision computation for the majority of blocks.
We pursue FGMP quantization where the majority of LLM weights and activations are quantized to low precision (e.g. FP4), with only a small percentage of values kept in higher precision (e.g. FP8).
As outlined in Figure \ref{fig:main}, we quantize the weights and activations at block granularity.
We use \textit{microscaling}, which defines a fine-grained scaling factor as metadata for each block of data, to reduce quantization error for the blocks that are quantized to low precision.
Microscaling has been shown to enable more accurate low-precision quantization \cite{dai2021vs,rouhani2023microscaling} with minimal hardware overhead.
Specifically, we leverage NVFP4 \cite{nvidia:NVFP4} as our low-precision datatype in order to accurately quantize each block to reduced precision.
Additionally, we propose hardware augmentations to support FGMP at block granularity while performing the majority of computation using low-precision datapaths.
Figure \ref{fig:ppl-vs-compression} highlights how our methodology achieves reduced perplexity degradation for the same compression ratio relative to existing work on weight/activation quantization.
As highlighted in Section \ref{sec:results-hardware}, utilizing low-precision datapaths also provides distinct energy savings, with NVFP4 datapaths with microscaling consuming 33\% less energy than FP8.
Our prototype demonstrates the efficiency benefits of our FGMP methodology. 

Our work makes the following contributions:
\begin{enumerate}
    \item We develop a policy to determine which blocks of weights and activations need to be retained in higher precision. We leverage sensitivity information with respect to the final model output to determine which blocks can be quantized to reduced precision without degrading the accuracy of the network. 
    Importantly, we ensure that this saliency policy is computationally efficient and can be computed on-the-fly for activations, thereby allowing for fast mixed-precision activation quantization.
    \item  We also design a fine-grained weight clipping strategy to improve the representation capacity for the blocks that are retained in lower precision.
    We tune the scale factor for each low-precision block to minimize the impact of quantizing the block on the model's output.
    \item We co-design the hardware to accelerate FGMP quantization:
    i) We adapt a Vector Multiply-Accumulate (VMAC) datapath to allow for performing mixed-precision dot products at 1D block granularity.
    Our datapath design supports FGMP quantization for both weights and activations.
    ii) We also design a mixed-precision activation quantization unit to assign activation blocks to low or high precision on the fly based on runtime statistics, thereby allowing for dynamically identifying and preserving regions of the activations that are challenging to quantize in higher precision.
\end{enumerate}

By leveraging our FGMP quantization method with NVFP4 and FP8 as the low and high precision number formats, we are able to attain the efficiency benefits of low-precision quantization for LLM inference while maintaining accuracy.
For the Llama-2-7B model, we attain 14\% energy savings in dot product computations during inference with $<1\%$ accuracy degradation (measured as perplexity on Wikitext-103).
Figure \ref{fig:ppl-vs-compression} also demonstrates how our method acheives less perplexity degradation for a target compression ratio relative to prior methods.
Our approach also requires 30\% less parameter memory than serving the model in FP8, enabling the serving of a larger model with the same memory budget.
\section{Related Work}

This section outlines prior work relevant to FGMP quantization. 
We first provide an overview of prior work on quantizing both weights and activations in LLMs.
We then discuss previous efforts to leverage mixed-precision quantization to improve accuracy with low-precision quantization.
Finally, we include a discussion of existing hardware support for mixed-precision quantization.

\subsection{LLM Quantization}

Quantization is a popular model compression method to improve inference efficiency.
Most existing quantization strategies have employed either integer or floating-point quantization \cite{lin2023awq,frantar2022gptq}.
More complex non-uniform quantization schemes allow for more flexible quantizaton signpost placement compared to integer and floating-point quantization \cite{kim2023squeezellm}; however, since quantization signposts are no longer uniformly placed, quantized values must first be dequantized and computation must be performed in higher precision. 
Prior work has also leveraged microscaling in order to enable accurate low-precision quantization \cite{dai2021vs,rouhani2023microscaling}. 
Microscaling formats leverage a fine-grained scaling factor for each block, which better preserves the values in the block by adjusting the represented range.

Typical quantization methods for LLMs have performed either weight-only \cite{lin2023awq,frantar2022gptq} or weight-activation \cite{shao2023omniquant,ashkboos2023towards} quantization.
In the weight-only regime, prior methods have been able to maintain high accuracy while compressing the model to 4-bit quantization; however, these approaches cannot utilize standard low-precision compute datapaths, as they still represented activations using higher precision.
In the weight-activation regime, it is possible to leverage low-precision compute datapaths to attain higher energy efficiency.
However, existing methods have struggled to retain accuracy in this regime due to the challenges of accurately quantizing both weights and activations to low precision without retraining \cite{xiao2023smoothquant,shao2023omniquant,liu2024spinquant}.
In this work, we aim to perform post-training weight-activation quantization in order to exploit low-precision compute units, while maintaining accuracy through retaining a small portion of blocks in higher precision.

\subsection{Mixed-Precision Quantization}

Several works have noted the presence of distinct outliers in both weights and activations, which significantly increase quantization difficulty ~\cite{dettmers2022llm,dettmers2023spqr,kim2023squeezellm,cui2024cherry}. 
Previous works have split LLM weight matrices into a dense low-precision matrix and an unstructured sparse high-precision matrix, which is used to accurately store numerical outliers and highly sensitive values \cite{dettmers2023spqr,kim2023squeezellm,cui2024cherry}. 
Although these approaches achieved high accuracy with low memory consumption for weight quantization, they cannot be easily applied for activation quantization and they lead to efficiency overheads in use cases that are not memory bandwidth-bound.
\cite{huang2024slim} employed fine-grained mixed-precision quantization at the 1D block granularity for LLM weights by leveraging sensitivity information, demonstrating the advantages of exploiting sensitivity information when determining which blocks to retain in higher precision.
However, this work was restricted to weight-only quantization, and was therefore unable to leverage standard low-precision compute datapaths.

There have also been previous works on coarse-grained structured mixed precision, in which a fixed block of the matrix is kept in higher precision, and the remainder of the matrix is quantized to low precision.
Several prior works have assigned different precisions to different layers (aiming to preserve more sensitive layers in higher precision) \cite{dong2019hawq,dong2020hawq,wang2019haq, zandonati2022fit}. 
Other works have performed structured mixed-precision quantization by preserving a subset of input channels in higher precision.
\cite{dettmers2022llm} preserved large-magnitude activation input channels (and the corresponding input channels in weights) in high precision, while quantizing the remaining values. 
\cite{ashkboos2023towards,zhao2024atom} reordered input channels in order to group large-magnitude activation channels together at the end of the matrix, and then left a portion of the input channels at the end of the matrix in higher precision for both weights and activations.
These works were able to leverage low-precision datapaths for part of the computation since they quantize both weights and activations.
However, these methods were unable to adapt to the unstructured nature of important values in weights and activations.
In contrast, our fine-grained mixed-precision approach can adapt to unstructured sensitive values in both weights and activations, while still performing the majority of computation using low precision datapaths.

There have also been prior works which searched for the optimal mixed-precision configuration, either by using reinforcement learning \cite{wang2019haq} or by formulated the search as a Neural Architecture Search problem \cite{wu2018mixed}. 
However, existing automatic mixed precision search methods are challenging to apply for LLMs, as these methods require repeated evaluation of the network with different precision configurations for each layer, which can be prohibitively expensive. 
Additionally, these approaches have previously been applied for layer wise mixed precision. 
With mixed precision at the 1D block granularity, the search space of potential precision assignments is much greater than the layer wise precision assignment problem.
It is therefore infeasible to leverage existing automatic methods to search for the optimal fine-grained mixed precision assignment.

\subsection{Hardware Support for Mixed-Precision Quantization}

Several prior works have investigated methods for adding architecture support for representing numerical outliers in order to improve low-precision quantization.
GOBO \cite{zadeh2020gobo} combines outlier-aware quantization with non-uniform quantization for better low-precision representation. 
However, this approach requires separately loading outliers and dense values (as the sparse values are stored in a separate sparse matrix), as well as separate datapaths for outlier processing. 
OLIVE \cite{guo2023olive} employs an outlier-aware encoding, storing outlier values across multiple positions (sacrificing the neighboring dense values).
They repurposed one of the 16 representative 4-bit values to indicate whether an outlier is present in a neighboring value, which amounts to an overhead of 1 bit per 4 elements. 
This has the benefit of allowing for a structured data storage format, but it negatively impacts the representation of nearby values.
One concurrent work, MicroScopiQ \cite{ramachandran2024microscopiq}, also retains outlier values in higher precision by pruning out a portion of non-outlier values and repurposing the saved bits to represent the outliers in higher precision.
SPARK \cite{liu2024spark} employs a variable-precision encoding scheme where large and small magnitude values are encoded with different precisions, and a bit at the start of each element indicates which precision is stored in that location.
This requires per-element metadata (eg. a bit for each low-precision element) to mark whether that element is in low or high precision. 
Both OLIVE and SPARK \cite{guo2023olive,liu2024spark} employ decoder modules at the boundary of a systolic array in order to decode their compact data format into a format that can be fed into an array of processing elements.
While these works perform decoding at the per-element level, we instead implement FGMP at block granularity.
This dramatically reduces metadata overhead compared to the per-element approach by amortizing the cost of the metadata across all elements in a block (so for a block size of 16, it requires only a single bit per 16 elements) and only requires performing a comparison with a single bit per block rather than performing per-element comparisons. 
FGMP greatly improves the efficiency of the hardware implementation by amortizing control overhead and enabling the use of efficient vector multiply-accumulate operations, while still providing sufficiently fine granularity to maintain high accuracy.

\section{Fine-Grained Mixed-Precision Quantization}

In this work, we focus on joint weight / activation quantization, specifically targeting the linear layers in LLM matrices, where a weight matrix $\textbf{W}$ and activation matrix $\textbf{X}$ are multiplied in order to compute the output $\textbf{Y}$, i.e. $\textbf{Y} = \textbf{W} \times \textbf{X}$.\footnote{Throughout this work, we use capital boldface letters to refer to matrices and lowercase boldface letters to refer to 1D blocks.}
The model is composed of $L$ layers, and the weights and activations for layer $l$ will be denoted as $\textbf{W}^{(l)}$ and $\textbf{X}^{(l)}$, respectively.
The model is trained by optimizing in order to reduce its loss $\mathcal{L}$, which is a function of both the weights and the activations corresponding to a given input data, as well as the ground truth labels corresponding to this data:

\begin{equation}
\label{eq:wa}
    \mathcal{L}(\{\textbf{X}^{(l)}\}_{l=1}^L, \{\textbf{W}^{(l)}\}_{l=1}^L)
\end{equation}

By minimizing the increase in the loss due to perturbation from quantizing the weights and activations, we can thereby minimize the perturbation in the model output, which is desirable for maintaining accuracy.

When we quantize an activation or weight value $v$ to precision $p$ using quantization function $Q_p$, we will express the quantization error at precision $p$, $\Delta_p v$, as follows:

\begin{equation}
\label{eq:delta}
   \Delta_p v = Q_p(v) - v
\end{equation}

In order to perform FGMP quantization, we first develop a policy to determine which weights and activations need to be retained in higher precision to preserve model accuracy.
While our experimental results in this work leverage FP8 and NVFP4 as the higher and lower precision number formats, respectively, our precision assignment policy developed in this section is generalizable to any choice of high-precision and low-precision formats.
We calibrate offline for a target sensitivity level above which the block is preserved in higher precision, and we apply a single threshold across different layers to allocate more high-precision blocks to more sensitive layers.
Finally, we develop a fine-grained weight clipping approach to improve the quantized representation of low-precision blocks with microscaling.
Taken together, our methodology facilitates accurate fine-grained mixed-precision quantization with only a small portion of blocks retained in higher precision.

\subsection{Precision Assignment Policy}

Our precision assignment policy determines which blocks to keep in higher precision in order to avoid degrading the accuracy of the model.
We can avoid this accuracy degradation by minimizing the degradation in the model's loss $\mathcal{L}$ when the weights and activations are perturbed.
We first consider how the loss is affected by perturbation in a single weight or activation value $v$ at a given layer $l$.
The loss can be expressed using Taylor expansion as follows (neglecting higher-order terms), where the gradient $g$ is the derivative of the loss with respect to the corresponding value, $\frac{\partial{\mathcal{L}}}{\partial{v}}$ :

\begin{align}
\label{eq:loss}
     \mathcal{L}(Q(v)) &= \mathcal{L}(v) + g \times \Delta v + \dots
\end{align}

To minimize the expected degradation in model loss, we therefore study the minimization of the following expression:

\begin{align}
    \label{eq:objective}
    \expect{v}{\lvert \mathcal{L}(v) - \mathcal{L}(Q(v)) \rvert^2} &= \expect{v}{g^2 \times (\Delta v)^2 }\\
    \label{eq:objective-2}
     &= \expect{v}{g^2} \times \expect{v}{(\Delta v)^2 }
\end{align}

Equation \ref{eq:objective-2} holds under the assumption that $g$ and $\Delta v$ are independent, meaning that the impact of perturbation in each weight or activation value on the model loss is independent of the perturbations in other values \cite{pmlr-v70-sakr17a}.
The expression $\expect{v}{g^2} $ corresponds to the diagonal Fisher information matrix.
Empirically, this can be evaluated by averaging the squares of the gradients over a set of calibration data.
For weight matrices, we calculate this by averaging the squared gradients over a sample dataset $D$ (in our case, 512 samples of sequence length 512 from the Wikitext-103 training set).
Since activation values are dynamic during inference, we do not have access to their gradients at runtime.
We instead use the average of the square of the gradients for each input channel (computed offline on the calibration set) to estimate per-channel sensitivity for activations.
For the remainder of the section, we will assume that gradients have been calibrated, and consequently, expectations are implied.

Considering a one-dimensional block $\textbf{v}$ (either an activation or weight block in layer $l$) containing $N$ elements, we can express the impact $I_{\mathcal{L}}$ of quantizing the block $\textbf{v}$ on the model's loss as the sum of the sensitivity-weighted quantization error for each element $v_i$ in $\textbf{v}$ (ignoring layer indices):

\begin{equation}
\label{eq:sensitivity}
    I_{\mathcal{L}}(\textbf{v}) = \sum_{i=1}^N g^2_i (\Delta v_i)^2 
\end{equation}

The above metric is used as a ranking mechanism to identify the most sensitive blocks, which are then preferentially kept in higher precision.
The increase in error for value $v$ when it is quantized in low precision $p_l$ rather than high precision $p_h$ can be expressed as follows:

\begin{equation}
\label{eq:error_delta}
    \Delta_{p_h \rightarrow p_l} v = \Delta_{p_l} v - \Delta_{p_h} v
\end{equation}

The impact on the final model output when a block $\textbf{v}$ of size $N$ containing values $v_i$ is quantized in low precision rather than high precision is therefore:

\begin{equation}
\label{eq:error_term}
    I_{\mathcal{L}}'(\textbf{v})  =  \sum_{i=1}^N g_i^2 (\Delta_{p_h \rightarrow p_l} v_i)^2
\end{equation}

We use the impact score $I_{\mathcal{L}}'(\textbf{v})$ to identify the weight and activation blocks for which low-precision quantization has the biggest impact on the model's output, and we preferentially preserve these blocks in higher precision.

\subsection{Setting the Threshold for Precision Assignment}

In order to determine which blocks can be quantized to reduced precision, we need to determine the threshold for the importance score $I_{\mathcal{L}}'(\textbf{v})$, above which the block must be retained in higher precision.
One option would be to compute the threshold for a target ratio of high-precision to low-precision blocks dynamically for each layer.
However, sweeping over the activation tensor for layer $l$ online during inference to compute $I_{\mathcal{L}}'(\textbf{v})$ for each block and then computing the target threshold for that layer would lead to unacceptable inference overhead (since we would need to write out each output block from the previous operation in both precisions before we knew which precision was going to be used).  
It is therefore preferable to calibrate offline to determine the threshold for activations for a target mixed-precision ratio.
Namely, for a target mixed-precision ratio $R$, we can assign $R$\% of blocks to be in higher precision by setting the threshold $T_{local}$ to be the $R$-th percentile of the impact score $I_{\mathcal{L}}'(\textbf{v})$, computed over all $J_l$ blocks $\textbf{v}$ in a single weight or activation tensor $\textbf{W}^{(l)}$ or $\textbf{X}^{(l)}$:

\begin{align}
\label{eq:local}
     T_{local} &= P_{R}(\{I_{\mathcal{L}}'(\textbf{v}^{(j)})\}_{j=1}^{J_l})
\end{align}

where $P_R()$ denotes the $R$-th percentile.
Additionally, different layers in the network exhibit differing sensitivities in terms of their impacts on the model output. 
In order to allow our policy to adapt to varying sensitivities at different layers, instead of calibrating for a separate threshold for each layer, we instead set a single global threshold across the entire model (one for weights and one for activations).
Since our sensitivity-weighted policy for precision assignment estimates the impact that quantizing the block will have on the final model output, it is already normalized across different layers (meaning that we can use the same threshold across different layers in the network even if they have different average magnitudes).
This allows for preserving a different proportion of sensitive blocks for particular layers which have a greater or lesser impact on the final model loss. 
For a target mixed-precision ratio $R$, we therefore set threshold $T_{global}$ to be the $R$-th percentile of the impact score $I_{\mathcal{L}}'(\textbf{v})$, computed over all blocks across all weight or activation tensors $\textbf{W}^{(l)}$ or $\textbf{X}^{(l)}$ (across all layers):

\begin{align}
\label{eq:global}
     T_{global} &= P_{R}(\{\{I_{\mathcal{L}}'(\textbf{v}^{(j)})\}_{j=1}^{J_l}\}_{l=1}^{L})
\end{align}

Note that this threshold is computed separately for weights and for activations since the input activation size will vary at inference time.
Figure \ref{fig:layer-distn} in Section \ref{sec:results-ablations} visualizes how our global threshold policy allows for retaining a greater portion of blocks in high precision for particular layers.

\subsection{Fine-Grained Sensitivity-Weighted Clipping}

Improving accuracy with low-precision quantization can provide noticeable benefits even in the context of FGMP quantization, as it can allow us to retain a greater portion of blocks in low precision without compromising accuracy. 
One method to improve low-precision quantization is clipping, where we reduce the representable range of numbers in order to give us better resolution within this range.
In the context of low-precision quantization with microscaling, performing clipping by adjusting the per-block scale factors can provide significant benefits as it allows us to perform fine-grained adjustments to best represent the values in a given block.
To improve the representation of blocks that are quantized to reduced precision, we therefore pursue clipping with the per-block scale factors.
In particular, we develop a sensitivity-weighted approach for fine-grained clipping to accurately quantize low-precision blocks without perturbing the model loss.
For a given block $\textbf{v}$ with $N$ elements in a weight layer $\textbf{W}^{(l)}$, the objective is to minimize the squared quantization error in $\textbf{v}$ weighted by the square of the gradients $g$. 
We therefore want to choose the scaling factor $s$ which minimizes this objective:

\begin{equation}
\label{eq:sw-clip}
    \min_s {\sum_{i=1}^N g^2_{i} (\overset{s}{\Delta}  v_i)^2}
\end{equation}

where $\overset{s}{\Delta} v_i$ is the quantization error for element $v_i$ when the per-block scaling factor is $s$.
With the low-precision datatype used in this work, the per-block scale factors are restricted to FP8 values, and as such there are only a limited number of possible values for $s$.
We can therefore perform a brute-force search over possible values for $s$ in order to identify the scale factors which minimize the sensitivity-weighted quantization error.
Note that we only apply sensitivity-weighted clipping offline for the weight matrices, and we use dynamic-max clipping online for activations, as it is not possible to efficiently compute the optimal scale factors for activations online during inference.

\subsection{Additional Baseline Policies}

\label{sec:additional-baselines}

\begin{figure}[t]
\centering
\includegraphics[width=\linewidth]{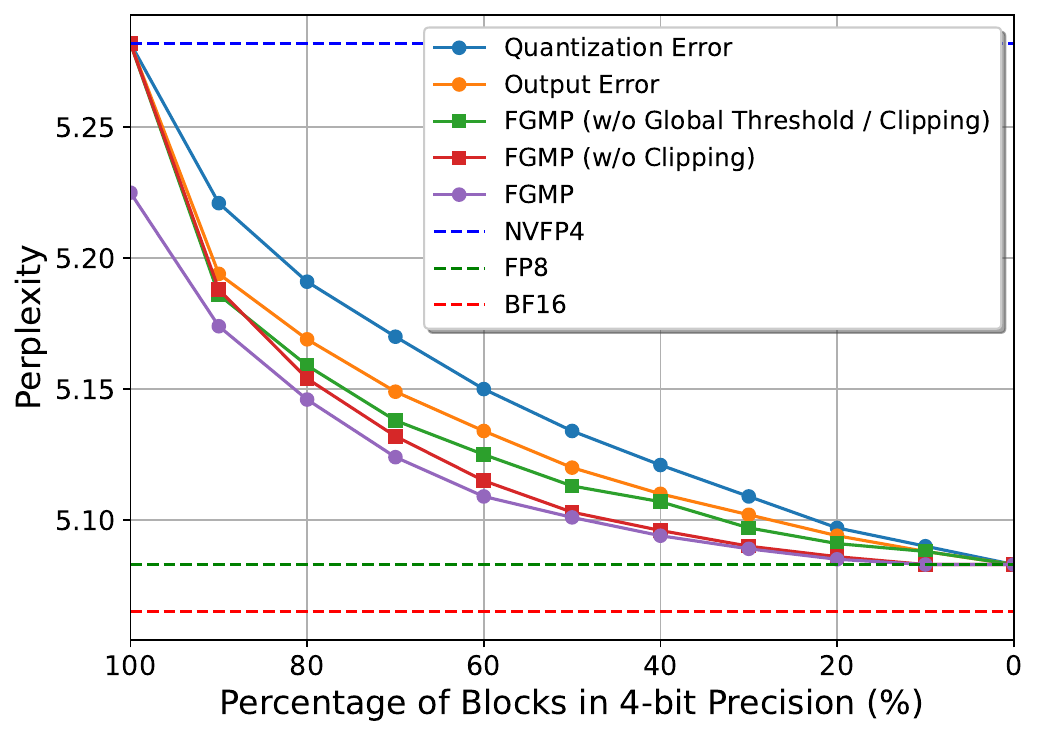}
\vspace{-5mm}
 \caption{
Perplexity evaluation on Wikitext-103 with the Llama-2-7B model (lower is better) with different percentages of blocks retained in FP8.
We provide an ablation of our approach against several baseline policies, and also highlight the perplexity benefits from using a single global threshold and from applying sensitivity-weighted clipping to the weights.
  }
  \label{fig:ppl-ablation}
\end{figure}

In order to justify our precision assignment policy, we compare with multiple potential baseline methods for selecting which blocks to retain in higher precision (the comparison is provided in Figure \ref{fig:ppl-ablation} in Section \ref{sec:results-ablations}). Our first baseline policy (referred to as ``Quantization Error'') determines which blocks to retain in higher precision using only the quantization error for each block within a single layer $l$. 
The error when a 1D block $\textbf{v}$ of size $N$ is quantized in low precision $p_l$ rather than high precision $p_h$ can be expressed as:

\begin{equation}
\label{eq:qerror}
    I_{QE}'(\textbf{v})  =  \sum_{i=1}^N  (\Delta_{p_h \rightarrow p_l} v_i)^2
\end{equation}

where $\Delta_{p_h \rightarrow p_l}$ gives the increase in error from quantizing element $v_i$ to low precision rather than high precision.

As a second baseline, we also include an ``Output Error'' policy, which minimizes the error at the layer output for layer $l$ in order to determine which blocks to retain in higher precision for that layer.
For this policy, we first calibrate for the average squared input channel magnitude across the other tensor $\textbf{Q}$.
For quantizing a block $\textbf{v}$ in weight matrix $\textbf{W}^{(l)}$, we compute the average squared magnitudes for the corresponding input channels in $\textbf{X}^{(l)}$, and for quantizing a block $\textbf{v}$ in activation matrix $\textbf{X}^{(l)}$ we compute the average squared magnitudes for the corresponding input channels in $\textbf{W}^{(l)}$.
We compute the average squared magnitudes for the input channels in $\textbf{X}^{(l)}$ statically using calibration data from the Wikitext-103 training set.
Using this policy, the error when a 1D block $\textbf{v}$ of size $N$ is quantized in low precision $p_l$ rather than high precision $p_h$ can be expressed as:

\begin{equation}
\label{eq:outerror}
    I_{OE}'(\textbf{v})  =  \sum_{i=1}^N \mathrm{avg}( Q_i^2 ) (\Delta_{p_h \rightarrow p_l} v_i)^2
\end{equation}

where $\mathrm{avg}( Q_i^2 )$ is the average squared input channel magnitude in the other tensor.
For both baseline policies, in order to set the mixed-precision ratio $R$, we set the threshold $T$ to be the $R$-th percentile of $I_{QE}'(\textbf{v})$ / $I_{OE}'(\textbf{v})$ computed dynamically over all blocks $\textbf{v}$ in a single weight or activation layer $\textbf{W}^{(l)}$ or $\textbf{X}^{(l)}$.

\begin{figure*}[t]
\centering
\includegraphics[width=0.95\linewidth]{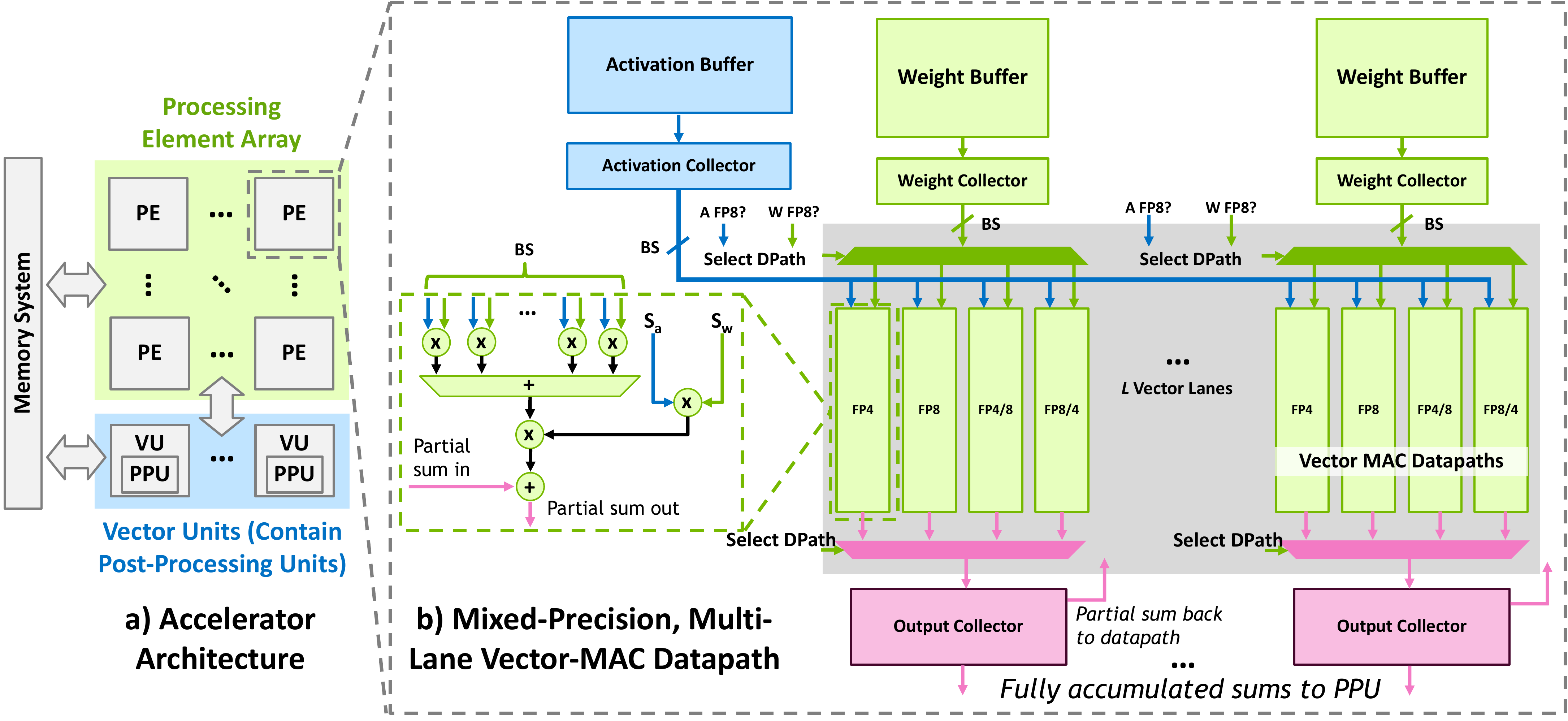}
\vspace*{-3mm}
 \caption{
a) High-level accelerator architecture, consisting of a PE array (each of which contains a VMAC-based datapath), as well as one or more vector units which contain our post-processing activation quantization unit (outlined in detail in Figure \ref{fig:main-hw2}).
b) Datapath support for FGMP quantization with four dot-product units per lane to perform FGMP VMAC operations.
The portion of the figure highlighted in gray is synthesized for hardware measurements. 
}
  \label{fig:main-hw1}
  \vspace{-3mm}
\end{figure*}

\begin{figure}[t]
\centering
\includegraphics[width=0.85\linewidth]{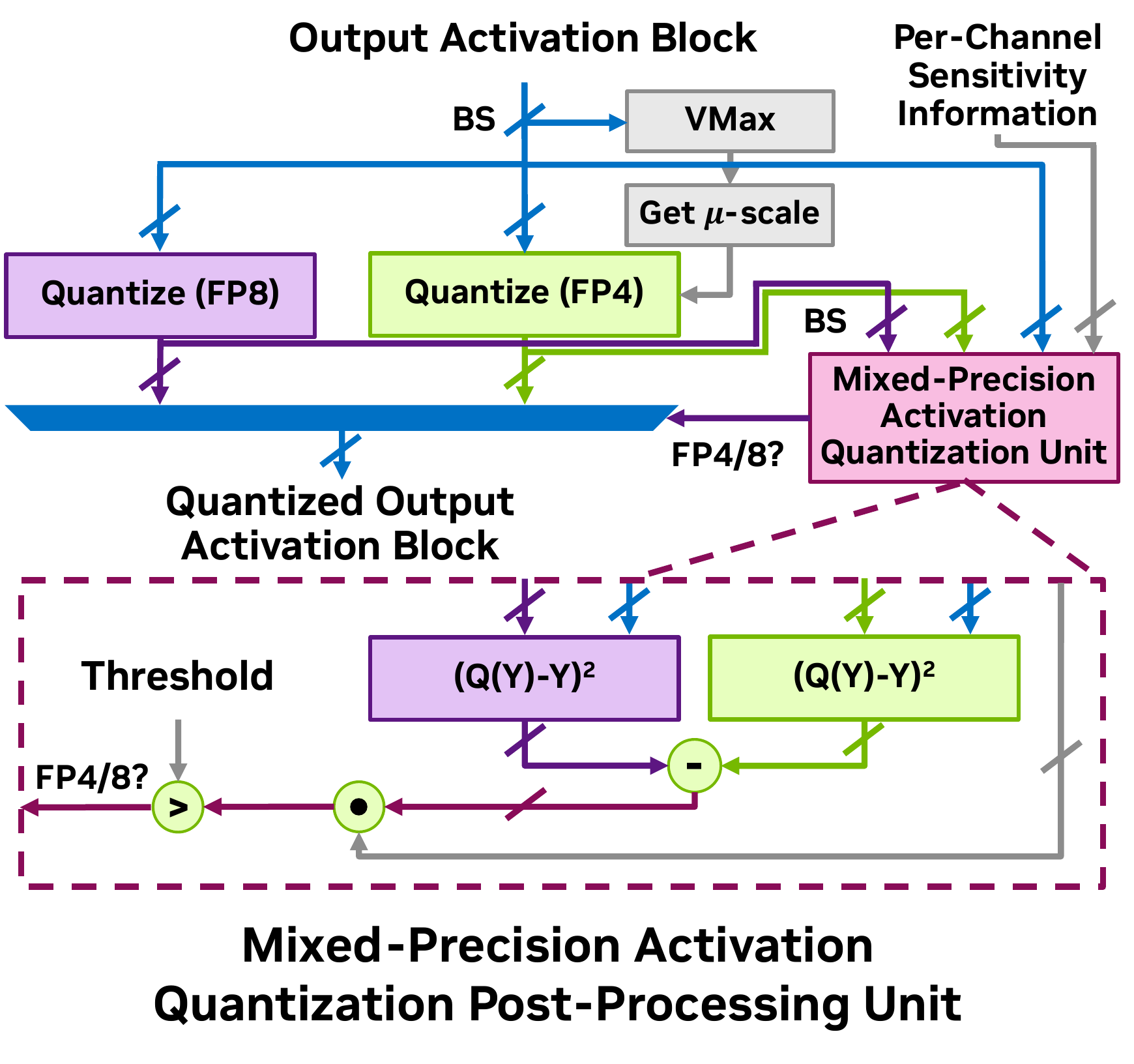}
\vspace*{-4mm}
 \caption{
A diagram of the post-processing unit for quantizing activations to mixed-precision online during inference.
  }
  \label{fig:main-hw2}
  
  \vspace{-6mm}
\end{figure}

\begin{figure*}[t]
\centering
\includegraphics[width=\linewidth]{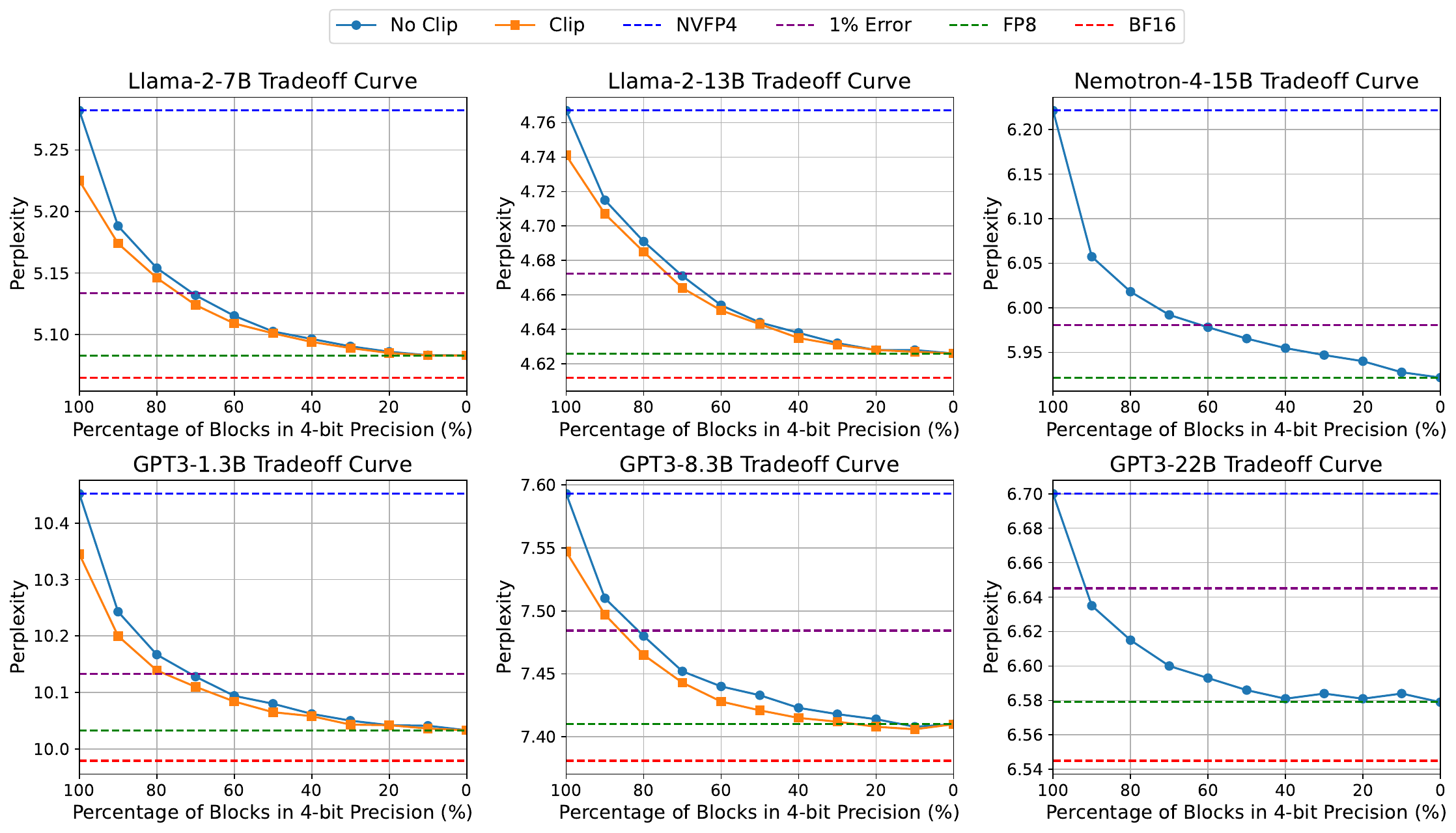}
\vspace*{-5mm}
 \caption{
Perplexity evaluation on Wikitext-103 for different models in the Llama-2, GPT3, and Nemotron model families. 
We report FGMP quantization results with and without our sensitivity-weighted weight clipping approach.
Note that for GPT3-22B and Nemotron-4-15B, we did not apply sensitivity-weighted clipping as it did not yield perplexity improvements, and for Nemotron-4-15B, we exclude the BF16 baseline as it exhibited worse perplexity than FP8.
  }
  \label{fig:wikitext103-ppl}
\end{figure*}

\section{Hardware Support for Fine-Grained Mixed-Precision Quantization}
\label{sec:hw}

Hardware support is required to use low-precision datapaths for the majority of computations when we leverage FGMP quantization.
Hardware support for quantizing activations to mixed-precision online during inference is also needed to realize the energy efficiency benefits of FGMP quantization, as these benefits would otherwise be negated by the overheads of performing mixed-precision quantization in software. Our proposed design (outlined in Figure \ref{fig:main-hw1}a) consists of an array of processing elements (PEs), each of which contains a Vector Multiply-Accumulate (VMAC) datapath consisting of multiple parallel lanes that each compute one VMAC per cycle \cite{venkatesan2019magnet}. 
The hardware augmentations required for supporting FGMP quantization for both weights and activations includes i) datapath modifications to support FGMP which are outlined in Figure \ref{fig:main-hw1}b, and ii) a post-processing unit (PPU) that efficiently quantizes activations to mixed precision during inference which is outlined in Figure \ref{fig:main-hw2}.
We use NVFP4 as our low precision datatype \cite{nvidia:NVFP4}, which uses microscaling to improve the quantization accuracy in low precision, as in \cite{dai2021vs,keller202217,rouhani2023microscaling}.
NVFP4 uses a block size of 16 and performs quantization using FP4 (E2M1) representation for the values with FP8 (E4M3) microscaling factors.
Throughout the remainder of this work, when we refer to a block being quantized in FP4, this block is quantized using the NVFP4 datatype.
We set the microscaling block size and the FGMP block size ($BS$) equal to the VMAC vector length to improve hardware efficiency.
We use a single metadata bit alongside each block to indicate whether the block contains high or low precision values.

\subsection{Mixed-Precision Datapath}

Figure \ref{fig:main-hw1}b outlines the multi-lane mixed-precision VMAC datapath which is required for performing efficient mixed-precision computation.
The datapath contains FP4, FP8, and mixed FP4/8 dot-product units.
In a single cycle, the active dot-product unit in each lane (determined by inspecting the metadata bit associated with each weight and activation block) computes $BS$-wide dot products across $L$ parallel lanes and adds this FP32 partial sum to the previously accumulated partial sum, thereby performing $2 \cdot BS \cdot L$ operations per cycle.
The dot-product units with at least one FP4 input also perform the scale factor multiplication before adding the partial sum to the accumulated value.
The remaining three dot-product units in each lane are inactive and are clock-gated or data-gated appropriately to minimize switching power.
When computing a matrix multiplication between a weight tile $\textbf{A}$ and an activation tile $\textbf{B}$, we assume a weight-stationary dataflow in which $\textbf{A}$ is held stationary and blocks in $\textbf{B}$ are streamed in one per cycle and broadcast across all lanes.
Note that we maintain the same math throughput per cycle regardless of the precision of the weights and inputs being consumed, which improves energy efficiency by maximizing temporal reuse of $\textbf{A}$ and spatial reuse of $\textbf{B}$ while avoiding complicated control logic.
Our proposed design contains independent FP4$\times$FP8 and FP8$\times$FP4 dot-product units in each lane so that the $A$ input can be consistently held constant during the innermost temporal loop, improving energy efficiency at the cost of area overhead.
Note that datapath components could be shared to reduce area; however, this would come at the cost of increased energy consumption.
Using shared hardware for all four dot-product combinations would obviate any energy gains of FGMP.
Sharing hardware between two or three dot-product formats could still realize some (reduced) energy savings.

\subsection{Post-Processing Activation Quantization Unit}

Figure \ref{fig:main-hw2} provides a diagram of our mixed-precision activation quantization unit, which is a post-processing unit that quantizes output activation blocks before writing them out to memory.
In a typical single-precision datapath, values are locally accumulated in high precision (e.g., FP32) before being sent to a PPU for scaling and quantization to the (single) activation data format.
In contrast, our mixed-precision PPU must dynamically determine whether to quantize the accumulated values to FP8 or FP4 by using the average per-input channel Fisher information to compute the sensitivity-weighted sum of the quantization error for each output block.
The activation block is then written out as either FP4 or FP8, depending on whether the sum is greater than the configured threshold.
The threshold is calibrated offline for a given model and is fixed throughout inference.
Note that our approach for leveraging per-input channel metadata for weighting the parameters in a block in order to determine the block's importance is agnostic to how this metadata is computed.
For example, this approach could also be used to run the baseline ``Output Error'' configuration described in Section \ref{sec:additional-baselines}, since it also uses statically calibrated per-input channel metadata to weight the values in each block when determining importance.

\subsection{Experimental Approach}

We prototype our hardware implementation using SystemC and the Siemens Catapult High-Level Synthesis tool to generate RTL. 
RTL simulations are performed using Synopsys VCS with synthetic data sampled from an appropriately scaled Gaussian random distribution, where the proportion of weight and activation blocks in FP4 and FP8 are independently adjusted to generate representative input stimulus.
We leverage the Synopsys Fusion Compiler for synthesis targeting a \qty{1}{\giga\hertz} clock period in a \qty{5}{\nano\meter} process, and we replay the test stimulus on the gate netlist to gather activity factors for power measurements using Synopsys PrimePower (TT \qty{0.67}{\volt} corner).
All measurements use a prototype design with $L=16$ and $BS=16$, as in \cite{dai2021vs}.

\begin{figure*}[t]
\centering
\includegraphics[width=0.85\linewidth]{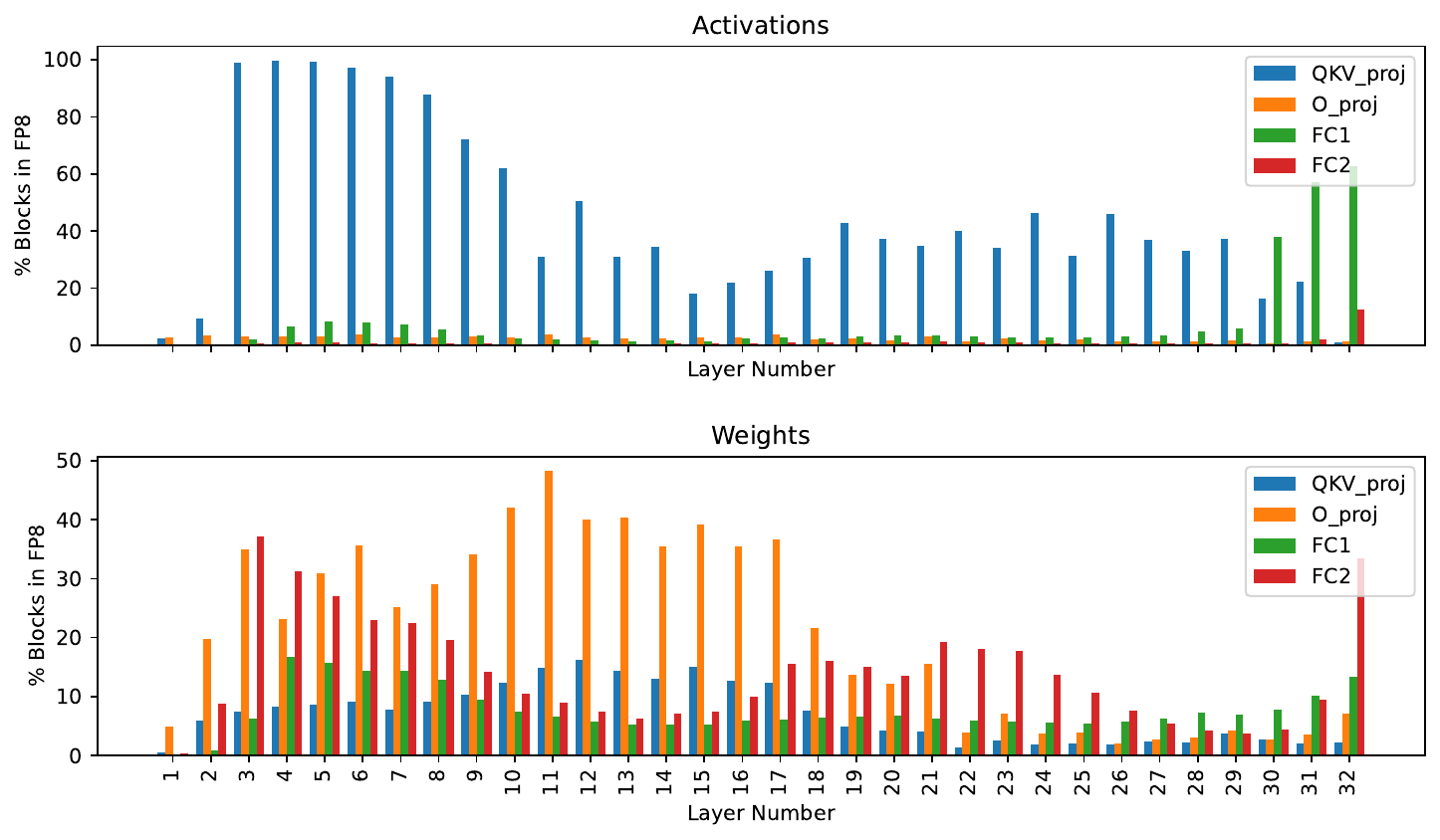}
\vspace*{-3mm}
 \caption{
Percentage of blocks retained in FP8 when using FGMP with sensitivity-weighted clipping targeting 90\% FP4 / 10\% FP8 for the Llama-2-7B model. 
We show the percentage of blocks in FP8 for all 32 QKV projection (``QKV\_proj''), Output projection (``O\_proj''), Fully Connected 1  (``FC1''), and Fully Connected 2 (``FC2'') layers. 
}
  \label{fig:layer-distn}
  \vspace{-3mm}
\end{figure*}

In order to collect energy estimates with a realistic workload, we need to leverage relevant estimates for the proportion of FP4 and FP8 blocks in different layers and for varying global threshold settings.
As outlined in Figure \ref{fig:layer-distn} in Section \ref{sec:results-ablations}, different layers have very different distributions of FP4 versus FP8 blocks, but measuring power for all possible layer configurations is runtime-intractable.
In order to collect realistic data, we first profiled the portion of FP8 and FP4 blocks across different layers for both weights and activations.
We collected this data for different global ratios of FP8 to FP4 blocks (from 90\% to 10\%).
We then treated each configuration as a set of features, normalized each feature, and performed K-means clustering to compute 100 representative configurations across all layers.  
After running power analysis on small kernels derived from each of these 100 configurations, we scaled up the results to match the shapes of each of the corresponding layers in order to estimate total energy consumption.

\begin{table}[t!]
\caption{
Perplexity on Wikitext-103 when applying sensitivity-weighted weight clipping ``SW-Clip") for Llama-2-7B and Llama-2-13B.
We report results for weight-only FP4 quantization with microscaling, both with and without sensitivity-weighted clipping.
All configurations use BF16 for activations.}
\label{tab:swclip}
\centering{
\scriptsize
\begin{tabular}{c|c|c}
 \toprule
    {\textbf{Weight Precision}} & {\textbf{Llama-2-7B}} & {\textbf{Llama-2-13B}} \\
    \midrule
   BF16   & 5.06 & 4.61 \\
   FP4    & 5.18 & 4.69 \\
   \hc \textbf{FP4 (w/ SW-Clip)}  & \textbf{5.13} & \textbf{4.67} \\
\bottomrule
\end{tabular}
\vspace{-4mm} 
 }
\end{table}

\begin{table}[t!]
\caption{
Average 5-shot evaluation results on MMLU for different models in the Llama-2, GPT3, and Nemotron model families. 
}
\label{tab:mmlu}
\centering{
\scriptsize
 \begin{tabular}{c|c|c|c|c|c|c}
     \toprule
        \multirow{2}{*}{\textbf{Model}} & \multicolumn{2}{c|}{\textbf{Llama-2}} & \multicolumn{3}{c|}{\textbf{GPT3}} & \textbf{Nemotron-4} \\
    & \textbf{7B} & \textbf{13B} & \textbf{1.3B} & \textbf{8.3B} & \textbf{22B} & \textbf{15B} \\
        \midrule
      BF16   & 0.458 & 0.551 & 0.247 & 0.252 & 0.387 & 0.643 \\
      FP8    & 0.462 & 0.550 & 0.249 & 0.248 & 0.383 & 0.647 \\
      FP4    & 0.430 & 0.533 & 0.258 & 0.258 & 0.371 & 0.634 \\
        \hc \textbf{90\% FP4} & \textbf{0.448} & \textbf{0.543} & \textbf{0.252} & \textbf{0.253} & \textbf{0.382} & \textbf{0.637} \\
        \hc \textbf{70\% FP4} & \textbf{0.451} & \textbf{0.547} & \textbf{0.256} & \textbf{0.251} & \textbf{0.382} & \textbf{0.642} \\
    \bottomrule
    \end{tabular}
\vspace{-3mm} 
 }
\end{table}

\begin{table}[t!]
\caption{Results on a selection of lm-eval-harness downstream evaluation tasks for different models in the Llama-2, GPT3, and Nemotron model families.
}
\label{tab:lmeval}
\centering{
\footnotesize{
\setlength{\tabcolsep}{3.5pt}{
    
    \centering
    \scriptsize{

    \vspace{-1mm}

    
    \begin{tabular}{c}
    \begin{tabular}{c|c|c|c|c|c|c}
     \toprule
        \multicolumn{7}{c}{\textbf{Llama-2-7B}} \\
        \midrule
    {\textbf{Precision}} & {\textbf{RACE}} & {\textbf{Hellaswag}} & {\textbf{PIQA}} & {\textbf{Winogrande}} & {\textbf{BoolQ}} & {\textbf{Average}} \\
        \midrule
        BF16   &  0.440 & 0.570 & 0.780 & 0.696 & 0.795 & 0.656 \\
        FP8    & 0.435 & 0.570 & 0.780 & 0.680 & 0.791 & 0.651 \\
        FP4    & 0.429 & 0.561 & 0.768 & 0.680 & 0.773 & 0.642 \\
        \hc \textbf{90\% FP4} & \textbf{0.427} & \textbf{0.564} & \textbf{0.777} & \textbf{0.684} & \textbf{0.782} & \textbf{0.647} \\
        \hc \textbf{70\% FP4} & 
        \textbf{0.450} & \textbf{0.569} & \textbf{0.777} & \textbf{0.695} & \textbf{0.788} & \textbf{0.656} \\
    \bottomrule
    \end{tabular}
    \end{tabular}
    \\
    \vspace{1mm} 
    
    \begin{tabular}{c}
    \begin{tabular}{c|c|c|c|c|c|c}
     \toprule
        \multicolumn{7}{c}{\textbf{Llama-2-13B}} \\
        \midrule
        {\textbf{Precision}} & {\textbf{RACE}} & {\textbf{Hellaswag}} & {\textbf{PIQA}} & {\textbf{Winogrande}} & {\textbf{BoolQ}} & {\textbf{Average}} \\
        \midrule
        BF16   & 0.448 & 0.602 & 0.797 & 0.723 & 0.822 & 0.678 \\
        FP8    & 0.450 & 0.603 & 0.792 & 0.721 & 0.824 & 0.678 \\
        FP4    & 0.440 & 0.595 & 0.785 & 0.715 & 0.809 & 0.669 \\
        \hc \textbf{90\% FP4} & \textbf{0.446} & \textbf{0.596} & \textbf{0.787} & \textbf{0.724} & \textbf{0.817} & \textbf{0.674} \\
        \hc \textbf{70\% FP4} & 
        \textbf{0.445} & \textbf{0.600} & \textbf{0.792} & \textbf{0.725} & \textbf{0.815} & \textbf{0.675} \\
    \bottomrule
    \end{tabular}
    \end{tabular}
    \\
    \vspace{1mm} 
    
    \begin{tabular}{c}
    \begin{tabular}{c|c|c|c|c|c|c}
     \toprule
        \multicolumn{7}{c}{\textbf{GPT3-1.3B}} \\
        \midrule
        {\textbf{Precision}} & {\textbf{RACE}} & {\textbf{Hellaswag}} & {\textbf{PIQA}} & {\textbf{Winogrande}} & {\textbf{BoolQ}} & {\textbf{Average}} \\
        \midrule
        BF16   & 0.363 & 0.439 & 0.742 & 0.591 & 0.648 & 0.557 \\
        FP8    & 0.365 & 0.439 & 0.737 & 0.580 & 0.642 & 0.552 \\
        FP4    & 0.359 & 0.429 & 0.731 & 0.577 & 0.642 & 0.548 \\
        \hc \textbf{90\% FP4} & \textbf{0.362} & \textbf{0.433} & \textbf{0.727} & \textbf{0.590} & \textbf{0.644} & \textbf{0.551} \\
        \hc \textbf{70\% FP4} & 
        \textbf{0.360} & \textbf{0.435} & \textbf{0.739} & \textbf{0.594} & \textbf{0.637} & \textbf{0.553} \\
    \bottomrule
    \end{tabular}
    \end{tabular}
    \\
    \vspace{1mm} 
    
    \begin{tabular}{c}
    \begin{tabular}{c|c|c|c|c|c|c}
     \toprule
        \multicolumn{7}{c}{\textbf{GPT3-8.3B}} \\
        \midrule
        {\textbf{Precision}} & {\textbf{RACE}} & {\textbf{Hellaswag}} & {\textbf{PIQA}} & {\textbf{Winogrande}} & {\textbf{BoolQ}} & {\textbf{Average}} \\
        \midrule
        BF16   & 0.415 & 0.546 & 0.780 & 0.680 & 0.689 & 0.622 \\
        FP8    & 0.416 & 0.546 & 0.774 & 0.672 & 0.691 & 0.620 \\
        FP4    & 0.415 & 0.533 & 0.774 & 0.665 & 0.683 & 0.614 \\
        \hc \textbf{90\% FP4} & \textbf{0.413} & \textbf{0.538} & \textbf{0.770} & \textbf{0.665} & \textbf{0.688} & \textbf{0.615} \\
        \hc \textbf{70\% FP4} & 
        \textbf{0.415} & \textbf{0.541} & \textbf{0.773} & \textbf{0.667} & \textbf{0.690} & \textbf{0.617}  \\
    \bottomrule
    \end{tabular}
    \end{tabular}
    \\
    \vspace{1mm} 
    
    \begin{tabular}{c}
    \begin{tabular}{c|c|c|c|c|c|c}
     \toprule
        \multicolumn{7}{c}{\textbf{GPT3-22B}} \\
        \midrule
        {\textbf{Precision}} & {\textbf{RACE}} & {\textbf{Hellaswag}} & {\textbf{PIQA}} & {\textbf{Winogrande}} & {\textbf{BoolQ}} & {\textbf{Average}} \\
        \midrule
        BF16   & 0.434 & 0.579 & 0.789 & 0.707 & 0.746 & 0.651 \\
        FP8    & 0.434 & 0.577 & 0.790 & 0.702 & 0.751 & 0.651 \\
        FP4    & 0.441 & 0.568 & 0.788 & 0.699 & 0.745 & 0.648 \\
        \hc \textbf{90\% FP4} & \textbf{0.441} & \textbf{0.573} & \textbf{0.784} & \textbf{0.707} & \textbf{0.740} & \textbf{0.649} \\
        \hc \textbf{70\% FP4} &
        \textbf{0.435} & \textbf{0.575} & \textbf{0.792} & \textbf{0.700} & \textbf{0.740} & \textbf{0.648}  \\
    \bottomrule
    \end{tabular}
    \end{tabular}
    \\
    \vspace{1mm} 
    
    \begin{tabular}{c}
    \begin{tabular}{c|c|c|c|c|c|c}
     \toprule
        \multicolumn{7}{c}{\textbf{Nemotron-4-15B}} \\
        \midrule
        {\textbf{Precision}} & {\textbf{RACE}} & {\textbf{Hellaswag}} & {\textbf{PIQA}} & {\textbf{Winogrande}} & {\textbf{BoolQ}} & {\textbf{Average}} \\
        \midrule
        BF16   & 0.471 & 0.620 & 0.812 & 0.756 & 0.787 & 0.689 \\
        FP8    & 0.482 & 0.625 & 0.813 & 0.757 & 0.736 & 0.683 \\
        FP4    & 0.457 & 0.615 & 0.791 & 0.743 & 0.689 & 0.659 \\
        \hc \textbf{90\% FP4} & \textbf{0.478} & \textbf{0.618} & \textbf{0.807} & \textbf{0.753} & \textbf{0.720} & \textbf{0.675} \\
        \hc \textbf{70\% FP4} & 
        \textbf{0.471} & \textbf{0.623} & \textbf{0.809} & \textbf{0.770} & \textbf{0.751} & \textbf{0.684}  \\
    \bottomrule
    \end{tabular}
    \end{tabular}
    \\
\vspace{-4mm} 
    }
     
     }
     }
     }
\end{table}

\begin{figure}[t]
\centering
\includegraphics[width=1\linewidth]{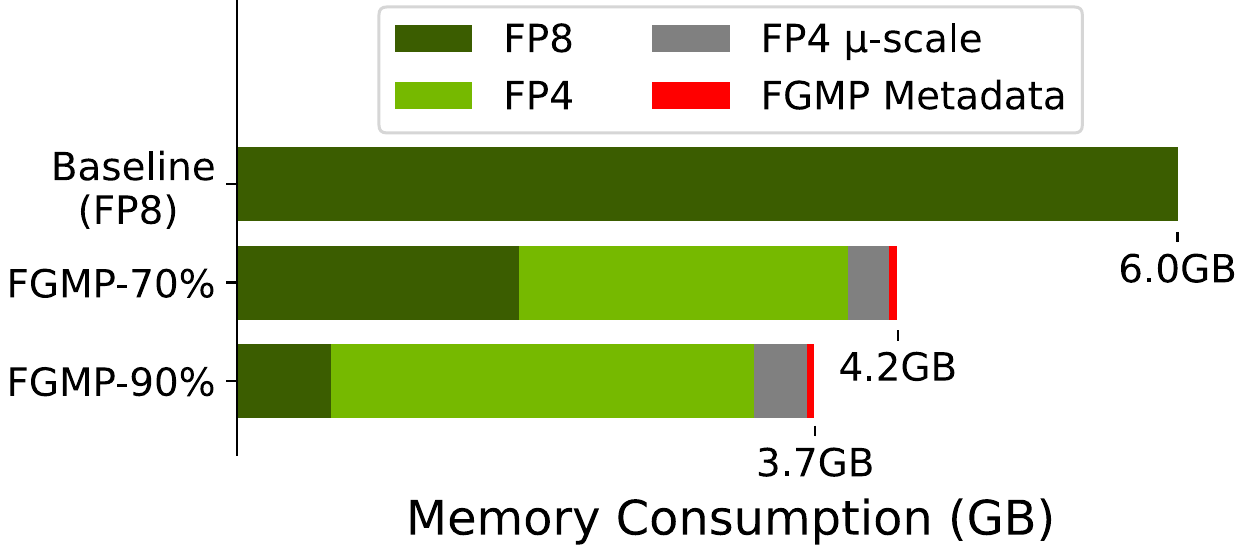}
\vspace*{-7mm}
 \caption{
Memory savings for the Llama-2-7B model weights when using FGMP (shown for configurations with 70\% and 90\% of blocks in FP4).
  }
  \label{fig:memory-savings}
\vspace{-5mm}
\end{figure}

\begin{figure}[t]
\centering
\includegraphics[width=0.8\linewidth]{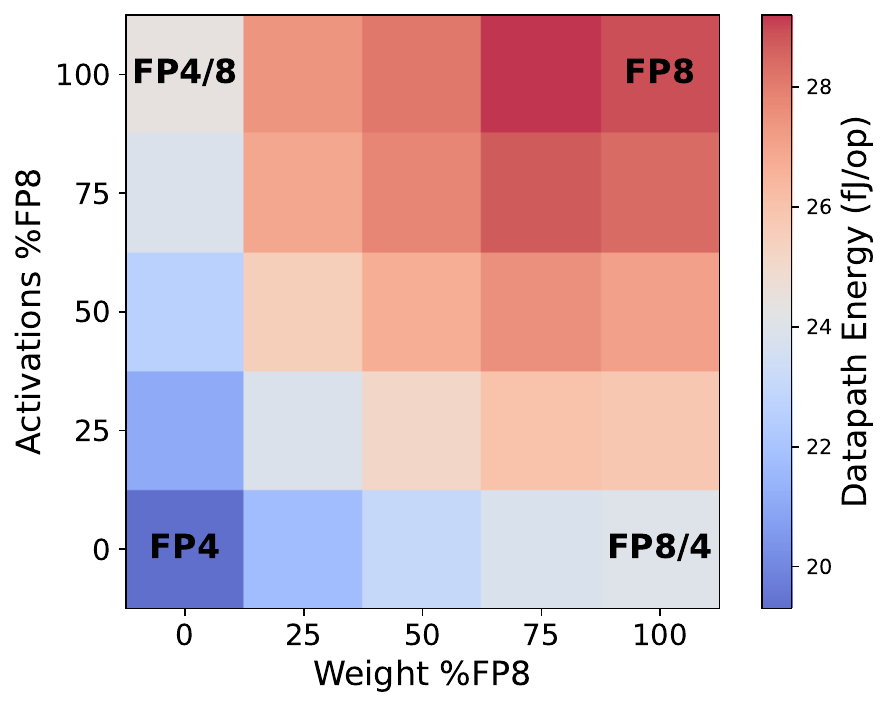}
\vspace*{-5mm}
 \caption{
Energy efficiency of the FGMP datapath processing different proportions of weights and activations in FP8.
The four labeled points show the energy efficiency when only a single dot-product unit is active over the entire test.
  }
  \label{fig:energy-colormap}
  \vspace{-5mm}
\end{figure}

\begin{figure}[t]
\centering
\includegraphics[width=\linewidth]{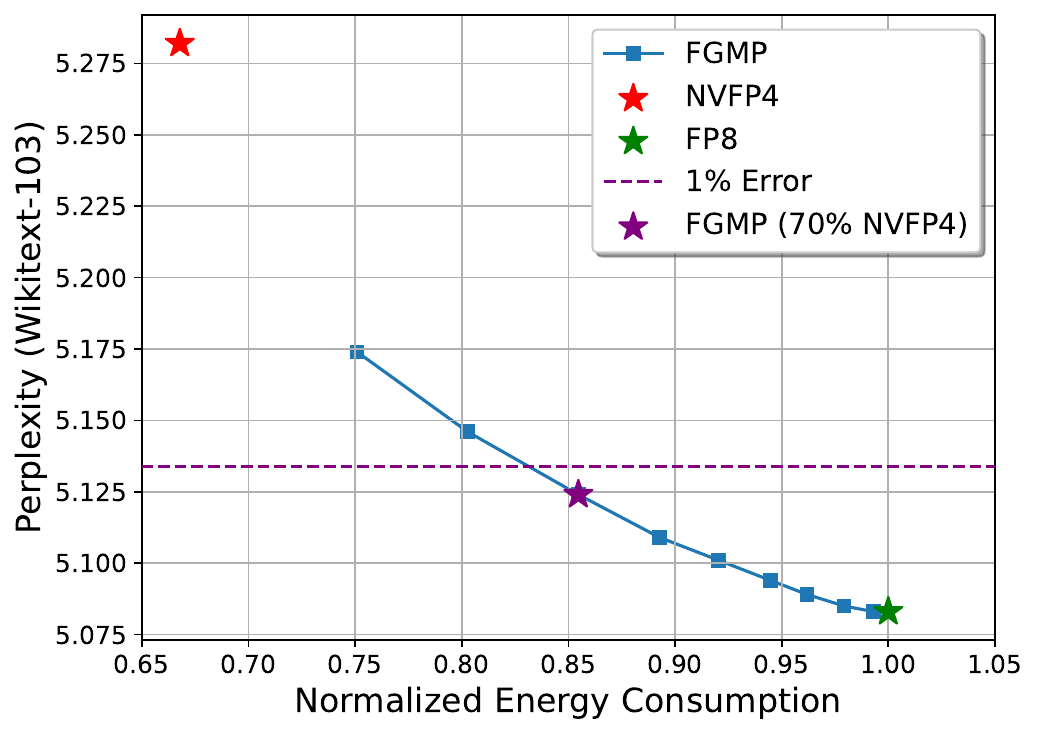}
\vspace*{-6mm}
 \caption{
Perplexity on Wikitext-103 versus normalized energy consumption using FGMP with sensitivity-weighted clipping with different percentages of blocks retained in FP8, using the Llama-2-7B model.
  }
  \label{fig:energy-placeholder}
  \vspace{-5mm}
\end{figure}

\section{Results}

We report a comprehensive accuracy evaluation of our quantization method and simulated energy measurements of our prototyped implementation in order to confirm the efficiency benefits of our approach. Sections \ref{sec:results-ppl} and \ref{sec:results-downstream} provide results for our methodology using perplexity as measured on Wikitext-103 and using downstream task evaluation, respectively.
Section \ref{sec:results-ablations} provides ablations for our precision assignment policy, as well as analysis for the distribution of sensitive blocks across different layers and the runtime for computing Fisher information for quantization.
Section \ref{sec:results-hardware} presents detailed evaluation of our hardware implementation.

\subsection{Perplexity Evaluation}
\label{sec:results-ppl}

We evaluate our proposed mixed-precision block assignment policy using models in the Llama-2, GPT3, and Nemotron families \cite{touvron2023llama,brown2020language,parmar2024nemotron}. 
We used FP8 without microscaling and NVFP4 as the higher and lower precision data formats.
Figure \ref{fig:wikitext103-ppl} shows the perplexity on Wikitext-103 versus the percentage of blocks that are kept in FP8, including results with and without our sensitivity-weighted clipping approach.
Our results show that with only a small percentage of blocks retained in FP8, we can attain significant accuracy improvements relative to quantizing all blocks to FP4.
These results also demonstrate the perplexity improvements from incorporating our sensitivity-weighted weight clipping approach (particularly when a higher proportion of blocks are quantized in FP4).
Figure \ref{fig:ppl-vs-compression} also provides comparisons against previous methods for the Llama-2-7B model, demonstrating how our method is able to attain improved perplexity for the same compression rate relative to prior methods.

\subsection{Downstream Task Evaluation}
\label{sec:results-downstream}

Tables \ref{tab:mmlu} and \ref{tab:lmeval} show the results of evaluating our methodology on downstream tasks. 
We compare FGMP against single-precision quantization on MMLU in Table \ref{tab:mmlu} as well as on a selection of tasks from {lm-eval-harness} (RACE, Hellaswag, PIQA, Winogrande, and BoolQ) in Table \ref{tab:lmeval} (with sensitivity-weighted weight clipping applied for Llama-2-7B/13B and GPT3-1.3B/8.3B).
On MMLU, we observe 58-89\% less accuracy degradation when we use FGMP with 70\% of blocks in FP4, relative to the degradation observed when going from FP8 to FP4 quantization.\footnote{Note that for GPT-3 1.3B/8.3B, the baseline accuracy on MMLU was close to 25\%, which is essentially random since each question is multiple choice with four possible answers.
This means that the results on these two models cannot be used to differentiate between configurations, and we therefore excluded these models from this calculation.}
On {lm-eval-harness}, we observe less than 0.4\% average accuracy degradation for all models when leveraging FGMP with 70\% of blocks in FP4 relative to FP8-only quantization.
Overall, our results demonstrate that our method helps retain downstream task performance even with the majority of blocks quantized to reduced precision.

\subsection{Ablation Studies}
\label{sec:results-ablations}

Figure \ref{fig:ppl-ablation} compares our FGMP approach with several other methods, including minimizing only the unweighted quantization error (``Quantization Error'') and minimizing the quantization error weighted by the average magnitude of the elements in the other tensor in order to minimize the error for the output of that particular layer (``Output Error'').
Additionally, we provide a comparison with using a per-layer dynamically-computed threshold in order to determine which blocks for that layer should be in FP4 or FP8 (``FGMP (w/o Global Threshold / Clipping)''), and with applying the global threshold but not applying our sensitivity-weighted clipping method (``FGMP (w/o Clipping)'').
These results demonstrate the superiority of our sensitivity-weighted precision assignment policy, which factors in the impact of each parameter on the final model output when assigning blocks to different precisions. 
These results also highlight the benefits of our global thresholding approach for maintaining high accuracy with the majority of blocks in low precision, as well as the perplexity benefits of our sensitivity-weighted weight clipping method.
Table \ref{tab:swclip} also provides additional evaluation for our sensitivity-weighted weight clipping approach, demonstrating the perplexity benefits in the weight-only quantization regime.

We also provide analysis for the portion of blocks retained in high precision across different layers.
Figure \ref{fig:layer-distn} shows the percentage of sensitive blocks that are retained in FP8 when using 90\% FP4 blocks and 10\% FP8 blocks (profiled using a sample of sequence length 4096 from the Wikitext-103 test set). 
This profiling demonstrates how our policy is able to adapt to differing sensitivities at different layers by allocating a larger or smaller portion of FP8 blocks to those layers (for example, retaining a greater portion of QKV projection layer activation blocks and Output projection layer weight blocks in FP8 at the early layers).

Additionally, to better understand the quantization efficiency of our method, we profiled the runtime for collecting the weight and activation Fisher information matrices.
For the Llama-2-7B model, we found that computing the Fisher information matrices (for 512 samples of sequence length 512) took less than 3 minutes on a single A100 GPU.
Note that this is a one-time cost for calibrating the model and can be performed ahead of inference time.
This demonstrates that our calibration procedure is lightweight and allows for quickly quantizing new models.

\subsection{Hardware Evaluation}
\label{sec:results-hardware}

\subsubsection{Memory Savings}

Figure \ref{fig:memory-savings} highlights the memory savings from leveraging our FGMP method for the Llama-2-7B model. 
We are able to attain 30\% and 39\% memory savings using our FGMP approach with 70\% and 90\% of blocks in FP4, respectively.
We also provide a breakdown of memory consumption, showing both the overhead of microscaling for the FP4 format as well as the overhead of the FGMP metadata (the per-block bit to distinguish FP4 and FP8 blocks). 

\subsubsection{Energy Analysis}

Figure \ref{fig:energy-colormap} shows the energy efficiency of our FGMP datapaths for different percentages of FP8 blocks in both weights and activations.
When evaluated only with stimulus restricted to a single data format (the labeled boxes in the figure), the NVFP4 datapath consumes 33\% less energy relative to the FP8 datapath, while the FP4/8 and FP8/4 (Weight / Activation) datapaths consume 16\% and 17\% less energy than the FP8 baseline, respectively.
The additional overheads of muxing between the different dot-product units at fine granularity imposes a small ``tax" to enable FGMP, such that ``mostly FP8" data costs slightly more energy than 100\% FP8 data.
However, this overhead is small compared to the energy savings of performing lower-precision arithmetic when most blocks are quantized to FP4.

We also measured the energy consumption of the mixed-precision activation quantization unit performing on-the-fly activation quantization under random stimulus.
The energy consumption of performing quantization on a single block is \qty{25.7}{\pico\joule}.
However, this operation need only be performed after reduction, so it is amortized over the dot product dimension of the input tensors, which is at least 4096 for the layers of the Llama-2-7B network.
Accordingly, this energy cost is just \qty{0.20}{\femto\joule\per{op}}, which is less than 1\% of the energy cost of the dot products themselves and is therefore negligible when considering system energy consumption. 

Figure \ref{fig:energy-placeholder} shows the perplexity on Wikitext103 when using our approach under different energy consumption budgets.
FP4 and FP8 baseline measurements are included for reference.
This data shows how our implementation is able to attain high accuracy with low energy consumption.
In particular, with less than 1\% accuracy degradation relative to the FP8 baseline, we are able to attain 14\% energy savings.
These results highlight the efficiency gains from leveraging fine-grained mixed-precision quantization in order to enable accurate low-precision LLM inference.

\subsubsection{Area Analysis}

\begin{table}[t!]
\caption{
Area breakdown for the FGMP datapath and post-processing unit (post-synthesis area in a 5nm process). Datapath areas are reported for 16 lanes, and all values assume a block size of 16. Note that the PPU area can be amortized across multiple PEs (since it is only invoked when writing out activations, which is performed infrequently).
Note that ``FP8/NVFP4'' refers to the datapath which supports FP8 weights and NVFP4 activations.
}
\label{tab:area}
\centering{
\vspace{-3mm} 
\scriptsize
\begin{tabular}{c|c}
 \toprule
    {Configuration} & {Area ($um^{2}$)} \\
   \midrule
   FP8 Datapath    & 2995  \\
   NVFP4 Datapath   & 1811  \\
   FP8/NVFP4 Datapath & 2669 \\ 
   NVFP4/FP8 Datapath & 2630  \\
   \hc FGMP Datapath   & 10356  \\
   \midrule
   FGMP PPU    & 8848 \\
\bottomrule
\end{tabular}
\vspace{-6mm} 
 }
\end{table}

Table \ref{tab:area} shows the area consumption of our custom design. 
The area overhead is 3.5$\times$ from a standalone FP8 datapath, or a 2.2$\times$ overhead from a datapath supporting only coarse-grained mixed precision in FP8 and FP4.
As noted in Section~\ref{sec:hw}, we chose a hardware design point that spends area to maximize energy efficiency; area overhead could be reduced by sharing datapath hardware.
Furthermore, it is important to note that datapath area is often a small portion of the area of a DNN accelerator. 
For example, the datapath is only 11\% of the area of a processing element in \cite{shao2021simba}. 
The processing elements are an even smaller portion of full system area, which tends to be memory-dominated.

Additionally, the PPU has an 85\% area overhead relative to the FGMP datapath with 16 lanes.
However, the PPU overhead can be amortized by increasing the number of lanes, or by sharing a PPU across multiple PEs. 
Given $P$ PEs with $L$ vector lanes per PE and $U$ PPUs, and assuming block size of 16, the time for the datapaths to process an $(M$ by $K) \times (K$ by $N)$ matrix (assuming a balanced pipeline) would be $\frac{M}{L}\cdot\frac{K}{16} \cdot\frac{N}{P}$ cycles, whereas the time for PPU processing would be $\frac{M}{16} \cdot \frac{N}{U}$ cycles. For a typical Llama-2-7B matrix multiplication with 4K context length ($4096$ by $4096 \times 4096$ by $4096$), a single PPU would be able to support up to 256 16-lane PEs without stalling; the area overhead of the PPU is therefore minimal when the cost is amortized across several PEs.
\section{Conclusion}

In this paper, we propose a post-training methodology for fine-grained mixed-precision quantization with both weights and activations. 
This approach allows us to retain the performance of the base model with only a small percentage of blocks retained in higher precision.
This is achieved by leveraging sensitivity information with respect to the final model output to determine which blocks should be preferentially retained in higher precision in order to preserve model accuracy.
Additionally, we develop a sensitivity-weighted clipping approach for weights that significantly improves accuracy and allows for a greater portion of blocks to be quantized to low precision with minimal accuracy loss.
In order to leverage the efficiency benefits of performing multiplications in reduced precision, we propose custom hardware support for FGMP quantization at the dot product level.
Our hardware implementation includes datapath support for performing dot products between two FP4 blocks, two FP8 blocks, or one FP4 and one FP8 block.
Our hardware implementation also includes a mixed-precision activation quantization unit that assigns activation blocks to low and high precision on-the-fly.
Our approach facilitates FGMP quantization, attaining the benefits of low-precision quantization while preserving the accuracy of inference with the high-precision values.

\bibliographystyle{ACM-Reference-Format}
\bibliography{sample-base}

\end{document}